\documentclass[12pt,preprint]{aastex}
\begin{document}
\title{Far Ultraviolet Spectral Analysis of the Prototype Nova-Like Variable VY Sculptoris
       from the High State to the Low State}

\author{Ryan T. Hamilton}
\affil{Dept. of Astronomy,
New Mexico State University,
Las Cruces, NM 88003,
email: rthamilt@nmsu.edu}

\author{Edward M. Sion}
\affil{Dept. of Astronomy \& Astrophysics,
Villanova University,
Villanova, PA 19085, 
e-mail: edward.sion@villanova.edu}

\begin{abstract}

The prototype nova-like variable VY Sculptoris was observed by the IUE during four different optical 
brightness states of the system. The FUV flux level from the highest state to the lowest state declines 
by a factor of 28. We have carried out model accretion disk and white dwarf atmosphere fitting to the 
spectra. The corresponding accretion rates range from $\dot{M}= 8 \times 10^{-9}$M$_{\sun}$ yr$^{-1}$ 
at the highest FUV flux level down to $\dot{M}= 1.9\times 10^{-10}$M$_{\sun}$ yr$^{-1}$ at the lowest flux level.
We report tentative evidence for the detection of the underlying white dwarf with $T_{\rm eff} = 45,000$K 
in the spectrum with the lowest flux level.
\end{abstract}

Subject Headings: keyword{Stars} Stars: cataclysmic variables, white dwarfs, Physical
Processes: accretion, accretion disks

\section{Introduction}

Cataclysmic variables (CVs) are short-period binaries in which a
late-type, Roche-lobe-filling main-sequence dwarf transfers gas through an
accretion disk onto a rotating, accretion-heated white dwarf (WD). The
nova-like variables are a subclass of CVs in which the mass-transfer rate
is high and the light of the system is dominated by a very bright
accretion disk. The spectra of nova-like variables resemble those of
classical novae (CNe) that have settled back to quiescence. However,
nova-like variables have never had a recorded CN outburst.  Hence their
evolutionary status is a mystery. They could be close to having their next
CN explosion, or they may have had an unrecorded explosion, possibly
hundreds or thousands of years ago. Adding to the mystery of nova-like
variables is that some of them (known as the VY Sculptoris stars after
their prototype) show the curious behavior of being in a high optical
brightness state most of the time, but then, for no apparent reason,
plummeting into a deep low-brightness state with little or no ongoing
accretion. Then, just as unpredictably, they return to the high-brightness
state. These drops are possibly related to cessation of mass transfer from
the K-M dwarf secondary star either by starspots positioned under L$_{1}$
(Livio and Pringle 1994) or irradiation feedback in which an inflated
outer disk can modulate the mass transfer from the secondary by blocking
irradiation of the hot inner accretion disk region(Wu et al. 1995).

VY Scl's brightest state has V = 12.9 and its low state has V = 18.5
(Warner 1987). Its orbital period is 0.1662 days but little else is known
(Hollander, A., Kraakman, H., van Paradijs, J.1993 and references
therein). Martinez-Pais et al (2000) found the white dwarf mass to be
$M_{(wd)} = 1.22\pm{0.22}M_{\sun}$, the mass of the secondary $M_{(2)} =
0.43\pm{0.13} M_{\sun}$ and the binary inclination $i = 30^{o}\pm{10} $ In
its low state, VY Scl shows emission lines similar to dwarf novae in
quiescence (Szkody 1987). A reddening correction of E(B-V) = 0.06 is given
by Bruch and Engle (1995). Unfortunately, the distance is unknown.

During the low states of VY Scl stars, it is possible to obtain nearly
uncontaminated spectra of the WD component because the accretion disk is
very faint and cool if present at all. Thus, through model-atmosphere
fitting during the low states, it is possible to determine temperatures,
gravities, and abundances for the WD.  This paper reports an analysis of
FUV archival spectra which sample several different brightness states of
the prototype, VY Scl itself. Among the key questions we seek to answer
are the following: What are the accretion rates during outburst and
quiescence? Is this rate of accretion consistent with the suppression of
the disk instability mechanism? How much flux does the white dwarf
contribute to the FUV?  How much flux does the accretion disk contribute
to the FUV? How hot is the white dwarf?

\section{Far Ultraviolet Spectroscopic Observations}

Far ultraviolet spectra of VY Scl were obtained over a period of four
years with the International Ultraviolet Explorer Spacecraft short
wavelength prime camera (SWP) through the large aperture at low dispersion
with a resolution of 5\AA. The SWP spectra covered the wavelength range
1170 \AA\ - 2000 \AA.  Fortuitously, the series of observations included
four spectra, SWP32594, SWP29755, SWP06123 and SWP21523, which covered a
variety of brightness states from the optical high brightness state when
the system was most luminous to intermediate brightness states down the
low optical brightness state when, in VY Scl nova-like variables, the
underlying white dwarf is thought to be exposed and the accretion disk had
greatly declined in brightness or gone away altogether. Due to large gaps
in the AAVSO observations archive for VY Scl, we cannot display the
placement of the observations on a light curve.  However, the AAVSO data
do indicate that at the time that SWP21523 was taken, the system was in a
deep low state with V < 15.8. Surprisingly, only one of the spectra,
SWP32594, has been discussed in the literature (LaDous 1991; Hoare and
Drew 1993). In Table 1, an observing log of the IUE archival spectra is
presented in which by column: (1) lists the SWP spectrum number, (2)  the
aperture diameter, (3) the exposure time in seconds, (4) the date and time
of the observation, (5) the continuum to background counts, and (6) the
brightness state of the system.

\begin{deluxetable}{lccccccrcl}  
 \tablecaption{IUE Observing Log
}
\tablenum{1}
\tablehead{                                   
\colhead{SWP}
&\colhead{t$_{exp}$}
&\colhead{Disp.}
&\colhead{Ap.}
&\multicolumn{2}{c}{Date of Observation}
&\colhead{F($\lambda$1350)}
&\colhead{C}
&\colhead{B}
&\colhead{State}
}
\startdata        
32594 &  2100&   LOW&    Lg&   1987-12-23 &12:51:00& $  5.5\times10^{-13}$&   144&   20 &  High State\\
29755  &  420 &  LOW&    Lg&   1986-11-26 &14:44:00&  $3.5\times10^{-13}$&    41&   17 &  High State\\
06123 &  2700 &  LOW&    Lg&   1979-08-09& 01:02:00&$  1.7\times10^{-13}$&   196 &  24 &  Intermediate\\
21523  & 7200 &  LOW&    Lg&   1983-11-12& 15:01:00&$  2.0\times10^{-14}$&    65&   30 &  Deep Low State\\
\hline
\enddata
\end{deluxetable}

The archival IUE NEWSIPS spectra were flux calibration-corrected using the
algorithm of Massa and Fitzpatrick (2000). Massa and Fitzpatrick (2000)
have shown that the absolute flux calibration of the NEWSIPS low
dispersion data was inconsistent with its reference model and subject to
time-dependent systematic effects, which together, amount to as much as
10-15\%. Therefore, in order to correct the data and optimize the
signal-to-noise, we used the IDL programs which apply Massa-Fitzpatrick
corrections to VY Scl's low dispersion IUE data.

The reddening of VY Scl was determined based upon all estimates listed in
the literature. The three principal sources of reddening values are the
compilations of la Dous (1991), Verbunt (1987) and Bruch \& Engel (1999).
The compilation of Bruch \& Engel listed E(B-V) = 0.06 for VY Scl.
Therefore the spectra were de-reddened with this color excess using the
IUERDAF routine UNRED. The brightness states of VY Scl at the time of the
IUE observations were assessed by comparison with the AAVSO light curve
data (visual magnitude versus Julian Date).

In figure 1, we display the four spectra together on the same flux scale.
The spectrum with highest flux level (the top spectrum), SWP32594 is
followed successively downward by two spectra of intermediate brightness
and finally the bottom-most spectrum taken during a low state of VY Scl.

The outburst spectrum is dominated by absorption lines including P Cygni
profile structure at C IV 1550, blue-shifted absorption at Si IV and
absorption features due to C II 1335, Si II 1260, and N IV 1718. In
SWP29755, the flux level has fallen to $3.5 \times 10^{-13}$ and the
absorption lines have weakened although N V 1240 and Si IV 1400 remain
prominent. In spectrum SWP06123, C IV emission is seen but, in general,
there is little convincing evidence for any other absorption features and
the flux at 1350\AA\ has declined to $1.7 \times 10^{-13}$. In the low
state spectrum, SWP21523, all regions where absorption appeared in
outburst and intermediate spectra have been replaced by emission lines and
there is an underlying continuum.  The strongest such emission lines are
Si IV (1400) and C IV (1550).

\section{Synthetic Spectral Fitting}

We adopted model accretion disks from the optically thick disk model grid
of Wade \& Hubeny (1998). In these accretion disk models, the innermost
disk radius, R$_{in}$, is fixed at a fractional white dwarf radius of $x =
R_{in}/R_{wd} = 1.05$. The outermost disk radius, R$_{out}$, was chosen so
that T$_{eff}(R_{out})$ is near 10,000K since disk annuli beyond this
point, which are cooler zones with larger radii, would provide only a very
small contribution to the mid and far UV disk flux, particularly the SWP
FUV bandpass. The mass transfer rate is assumed to be the same for all
radii. Thus, the run of disk temperature with radius is taken to be:

\begin{equation}
T_{eff}(r)= T_{s}x^{-3/4} (1 - x^{-1/2})^{1/4}
\end{equation}

where  $x = r/R_{wd}$
and $\sigma T_{s}^{4} =  3 G M_{wd}\dot{M}/8\pi R_{wd}^{3}$

Limb darkening of the disk is fully taken into account in the manner
described by Diaz et al. (1996) involving the Eddington-Barbier relation,
the increase of kinetic temperature with depth in the disk, and the
wavelength and temperature dependence of the Planck function.  The
boundary layer contribution to the model flux is not included. However,
the boundary layer is expected to contribute primarily in the extreme
ultraviolet below the Lyman limit.

Model spectra with solar abundances were created for high gravity stellar
atmospheres using TLUSTY (Hubeny 1988) and SYNSPEC (Hubeny \& Lanz 1995).
Using IUEFIT, a $\chi^{2}$ minimization routine, both $\chi^{2}$ values
and a scale factor were computed for each model.  The scale factor,
normalized to a kiloparsec, can be related to the white dwarf radius
through: $F_{\lambda(obs)} = 4 pi (R^{2}/d^{2})H_{\lambda(model)}$.

After masking emission lines in the spectra, we determined separately for
each spectrum, the best-fitting white dwarf-only models and the
best-fitting disk-only models using IUEFIT, a $\chi^{2}$ minimization
routine. A dense grid of model spectra with solar abundances was created
for high gravity stellar atmospheres using TLUSTY (Hubeny 1988) and
SYNSPEC (Hubeny \& Lanz 1995). We took a range of gravities in the fitting
from Log $g = 7.0 - 9.0$ in steps of 0.5. For the white dwarf radii, we
use the mass-radius relation from the evolutionary model grid of Matt Wood
(private communication) for C-O cores.

Taking the best-fitting white dwarf model and combining it with the
best-fitting disk model, we varied the accretion rate of the best-fitting
disk model by a small multiplicative factor in the range 0.1 to 10 using a
$\chi^{2}$ minimization routine called DISKFIT. Using this method the
best-fitting composite white dwarf plus disk model is determined based
upon the minimum $\chi^{2}$ value achieved and consistency of the scale
factor-derived distance with the adopted distance for each system.  The
scale factor, $S$, normalized to a kiloparsec and solar radius, can be
related to the white dwarf radius R through:  $F_{\lambda(obs)} = S
H_{\lambda(model)}$, where $S=4\pi R^2 d^{-2}$, and $d$ is the distance to
the source.

The results of our fitting are summarized as follows. The spectra with the
three highest flux levels are all dominated by an accretion disk. The best
fitting disk models to SWP32594, SWP29755 and SWP06123 are displayed in
figures 2, 3, and 4. In each case the accretion disk provides essentially
all of the FUV flux even as the flux level at its peak in SWP32594 is
lower by a factor of 3 in SWP06123.

In figure 2, the best-fitting disk model to SWP32594 corresponded to the
following parameters:  accretion rate $\dot{M}= 8 \times
10^{-9}$M$_{\sun}$ yr$^{-1}$, $M_{(wd)} = $1.0M$_{\sun}$, and an
inclination of $41^{o}$ with a $\chi^{2}$ = 5.96 The scale factor of this
best fit yields a distance to VY Scl of 620 parsecs.

At the lower flux level of SWP29755, the best-fitting disk model is
displayed in figure 3. This fit corresponded to the following parameters:
accretion rate $\dot{M}= 5\times 10^{-9}$M$_{\sun}$ yr$^{-1}$, $M_{(wd)} =
$1.0M$_{\sun}$, inclination of $41^{o}$ with a $\chi^{2}$ = 7.86. The
scale factor of this fit yielded a distance of 616 pc.
 
The best-fitting disk model to SWP06123 corresponded to the following
parameters: accretion rate $\dot{M}= 1.6\times 10^{-9}$M$_{\sun}$
yr$^{-1}$, $M_{(wd)} = $1.0M$_{\sun}$, and an inclination of $41^{o}$ with
$\chi^{2}$ = 5.74. The scale factor S yielded a distance to VY Scl of 536
parsecs.  This fit is seen in figure 4

The spectrum with the lowest flux level, SWP21523 was obtained during a
low state of VY Scl. The flux level at 1350A is lower by a factor of 28
than the spectrum in Figure 1 with the highest flux level. This spectrum
contains a very significant contribution (34\% of the flux) from a a hot
WD but with the flux of an accretion disk still accounting for the
majority (66\%) of the FUV flux. If this spectrum recorded the deepest low
state of VY Scl, then the presence of a strong disk component is a
departure from the totally white dwarf-dominated low state spectra of the
low state nova-like systems MV Lyra, TT Ari and DW UMa. Our best-fitting
white dwarf plus accretion disk combination fit to SWP21523 consisted of a
hot solar composition WD with $T_{\rm eff} = 45,000$K$\pm3000K$, log$g =
8.5$ with a $\chi^{2}$ = 3.38 and accretion disk model with accretion rate
$\dot{M}= 1.9\times 10^{-10}$M$_{\sun}$ yr$^{-1}$, $M_{(wd)} =
$1.0M$_{\sun}$, and an inclination of $41^{o}$ with $\chi^{2}$ = 3.38. The
scale factor S yielded a distance to VY Scl of 536 parsecs. This fit is
seen in figure 5.

It is important to explore the uncertainties in our estimated accretion
rates, $\dot{M}$. In order to assess the errors in $\dot{M}$ through a
formal
error analysis, it is necessary to know the errors in the individual input
parameters used to determine $\dot{M}$. These inexact parameters are the
white dwarf mass, the orbital inclination, the distance to the system and
the reddening. Each would produce errors in $\dot{M}$ and each parameter
has
its own uncertainties. Since each of these parameters affects either the
continuum slope or the flux level or both, they affect the accretion rate.
For example, Puebla et al. (2007) using similar disk models as ours but
with a more sophisticated four dimensional multiparameter optimization
fitting method, cite an error in E(B-V) of 0.02 to 0.05 producing errors
in $\dot{M}$ of 30 to 50\%.  For a sense of the uncertainty of the
accretion
disk fitting method used in this paper, the reader is referred to Winter
and Sion (2003) who carried out a formal error analysis with error
contours for $\dot{M}$.

\section{Conclusions}

Our multi-component white dwarf photosphere plus accretion disk model
analysis of the archival IUE spectra of VY Scl covered four different
optical brightness states of the system. We find strong evidence that the
accretion rate has declined from $\dot{M}= 8 \times 10^{-9}$M$_{\sun}$
yr$^{-1}$ at the highest FUV flux level down to $\dot{M}= 1.9\times
10^{-10}$M$_{\sun}$ yr$^{-1}$ during the lowest brightness state
spectroscopically recorded by the IUE. During this latter state of lowest
flux, the white dwarf contributes substantially to the FUV flux. This is
not unexpected for a nova-like low state since the underlying white dwarfs
in the nova-like systems TT Ari (Gaensicke et al. 2001), MV Lyr (Linnell
et al.2005), DW UMa (Knigge et al.2005) and V794 Aql (Godon et al.2007)
are also exposed. The temperature estimate of 45,000K we obtain for the
white dwarf in VY Scl must be regarded as preliminary. It is however
interesting that the temperature for the WD in VY Scl is within the range
of white dwarf temperatures measured for the other nova-like systems as
seen in Table 2.

\begin{deluxetable}{lccl}  
 \tablecaption{Temperatures of WDs in VY Sculptoris-Type Nova-like Variables
}
\tablenum{2}                                                                          
\tablehead{                                   
\colhead{Star}&\colhead{P$_{orb}$}&\colhead{d(pc)}&\colhead{Te}
}
\startdata        
V794 Aql& 0.2323 &250& > 47,000 K\\
TT Ari& 0.1375 &185 &40,000 K\\
MV Lyr &0.1329& 335& 47,000 K\\
DW UMa & 0.1375 & 830 & 46,000K\\
VY Scl &0.1662 &300 &45,000 K\\
\enddata\\
References: V794 Aql: Godon et al. (2007), ApJ, in press; TT Ari: Gaensicke et al. (2001); MV Lyr: Linnell et al. (2005); VY Scl: This paper.
\end{deluxetable}

Since dwarf nova behavior is not seen in VY Scl-type and UX UMa-type
nova-like variables, it is widely held that the supression of the disk
instability mechanism requires that the accretion rate be higher than the
critical mass transfer rate for dwarf nova outbursts. This threshhold
value as a function of white dwarf mass and orbital period is discussed in
detail by Shafter, Wheeler and Cannizzo (1986). Referring to Figure 2 in
Shafter et al.(1986), the accretion rates we estimated for VY Scl in its
high states lie above the dwarf nova instability line for white dwarf
masses from 0.4 to 1.0 Msun.

It is clear that higher quality FUV spectra are required to confirm the
tentative results herein for VY Scl.  Moreover, given the small size of
the sample of exposed white dwarfs in nova-like systems, it is
particularly important to catch more nova-like systems in their low states
for both ground-based optical and space observations.

\acknowledgments 

We thank an anonymous referee for helpful comments. This work was
supported by NSF grants AST05-07514, NNG04GE78G and by summer
undergraduate research support from the Delaware Space Grant Consortium.  
Some or all of the data presented in this paper were obtained from the
Multi mission Archive at the Space Telescope Science Institute (MAST).
STScI is operated by the Association of Universities for Research in
Astronomy, Inc., under NASA contract NAS5-26555. Support for MAST for
non-HST data is provided by the NASA Office of Space Science via grant
NAG5-7584 and by other grants and contracts.

Figure Captions

\figcaption{The four IUE spectra corresponding to four different brightness states
from the peak of outburst (top spectrum, highest flux level) down to 
a deep low state (bottom spectrum, lowest flux level).
Note the transition of the line spectrum as a function of brightness state
from the outburst spectrum where absorption lines dominate
down to the low state FUV spectrum when emission lines dominate the spectrum.}

\figcaption{The best-fit accretion disk
synthetic fluxes to the spectrum SWP32594 of VY Scl during its high state. The accretion
disk corresponds to $\dot{M} = 8\times 10^{-9} M_{\sun}$/yr, $i = 41
^{o}$, and M$_{wd} = 1.0 M_{\sun}$. The top solid curve is the
best-fitting combination, the dotted curve is the negligible contribution of the white dwarf 
alone and the dashed curve is the accretion disk synthetic spectrum alone.
In this fit, the white dwarf the accretion disk contributes 98\% of the
far UV flux and the white dwarf 2\% of the flux.}

\figcaption{The best-fit accretion disk
synthetic fluxes to the spectrum SWP29755 of VY Scl during its high state. The accretion
disk corresponds to $\dot{M} = 5\times 10^{-9} M_{\sun}$/yr, $i = 41
^{o}$, and M$_{wd} = 1.0 M_{\sun}$. The top solid curve is the
best-fitting combination, the dotted curve is the contribution of the white dwarf 
alone and the dashed curve is the accretion disk synthetic spectrum alone.
In this fit, the white dwarf the accretion disk contributes 98\% of the
far UV flux and the white dwarf 2\% of the flux.}

\figcaption{The best-fit accretion disk
synthetic fluxes to the spectrum SWP of VY Scl during its high state. The accretion
disk corresponds to $\dot{M} = 1.6\times 10^{-9} M_{\sun}$/yr, $i = 41
^{o}$, and M$_{wd} = 1.0 M_{\sun}$. The top solid curve is the
best-fitting combination, the dotted curve is the small contribution of the white dwarf 
alone and the dashed curve is the accretion disk synthetic spectrum alone.
In this fit, the white dwarf the accretion disk contributes 92\% of the
far UV flux and the white dwarf 8\% of the flux.}

\figcaption{The best-fit combination of white dwarf plus accretion disk
synthetic fluxes to the spectrum of VY Scl during its low state. The white
dwarf model has $T_{\rm eff} = 45,000$K, log $g =8.5$ and the accretion
disk corresponds to $\dot{M} = 1.9\times 10^{-10} M_{\sun}$/yr, $i = 41
^{o}$, and M$_{wd} = 1.0 M_{\sun}$. The top solid curve is the
best-fitting combination, the dotted curve is the white dwarf spectrum
alone and the dashed curve is the accretion disk synthetic spectrum alone.
In this fit, the accretion disk contributes 66\% of the
far UV flux and the white dwarf 34\% of the flux.}

%\clearpage

\plotone{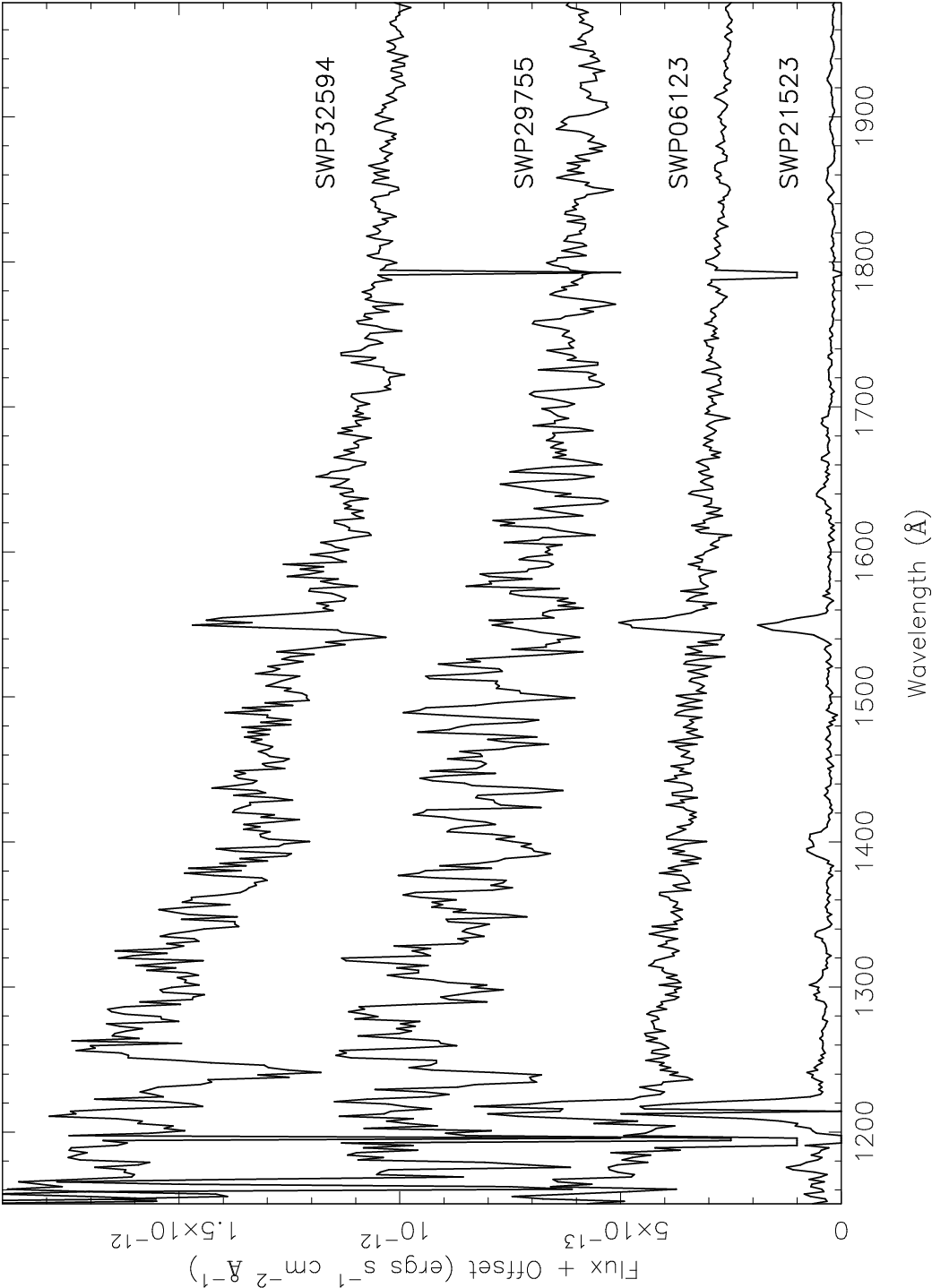}
%\clearpage

\plotone{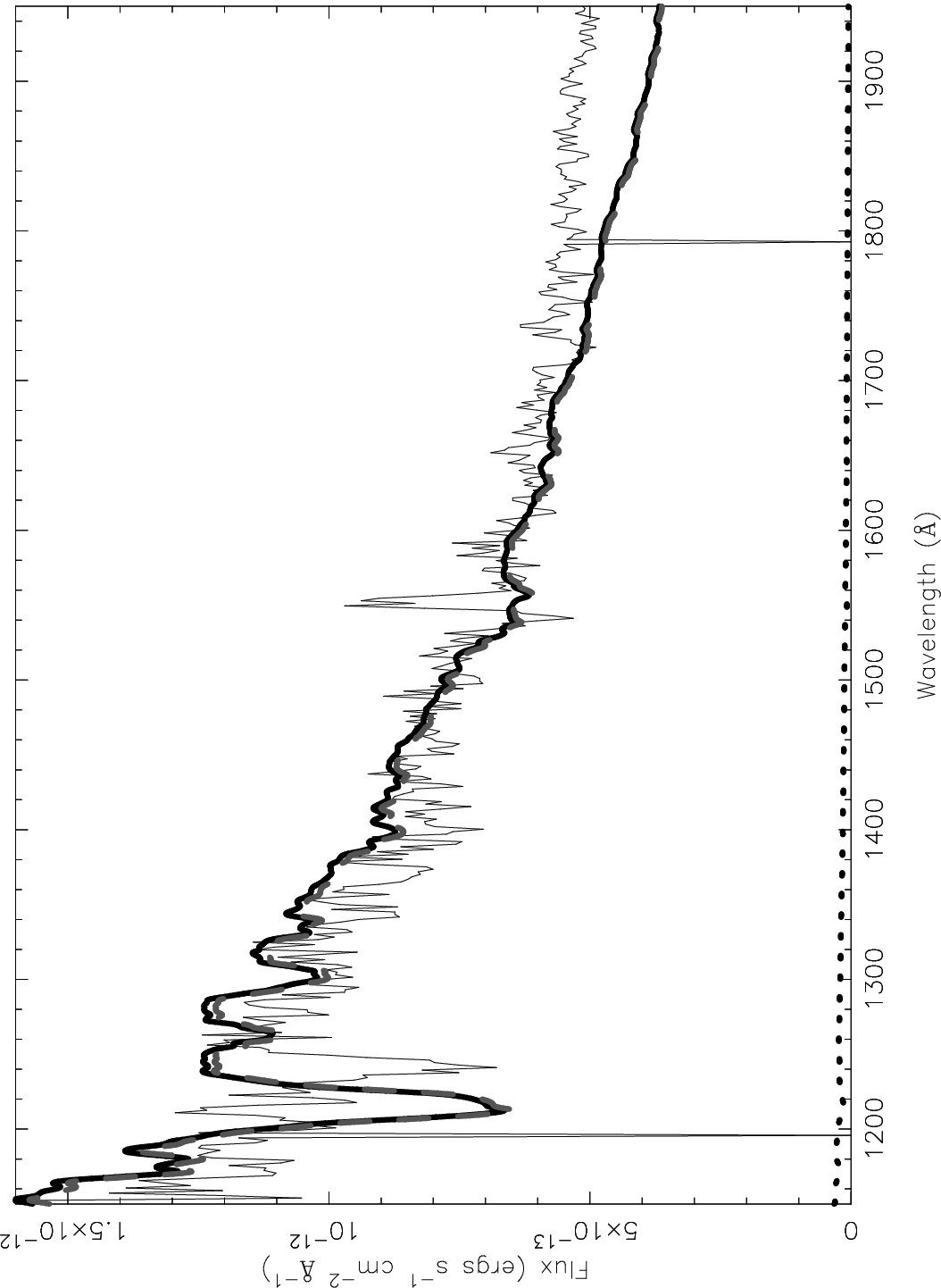}
%\clearpage

\plotone{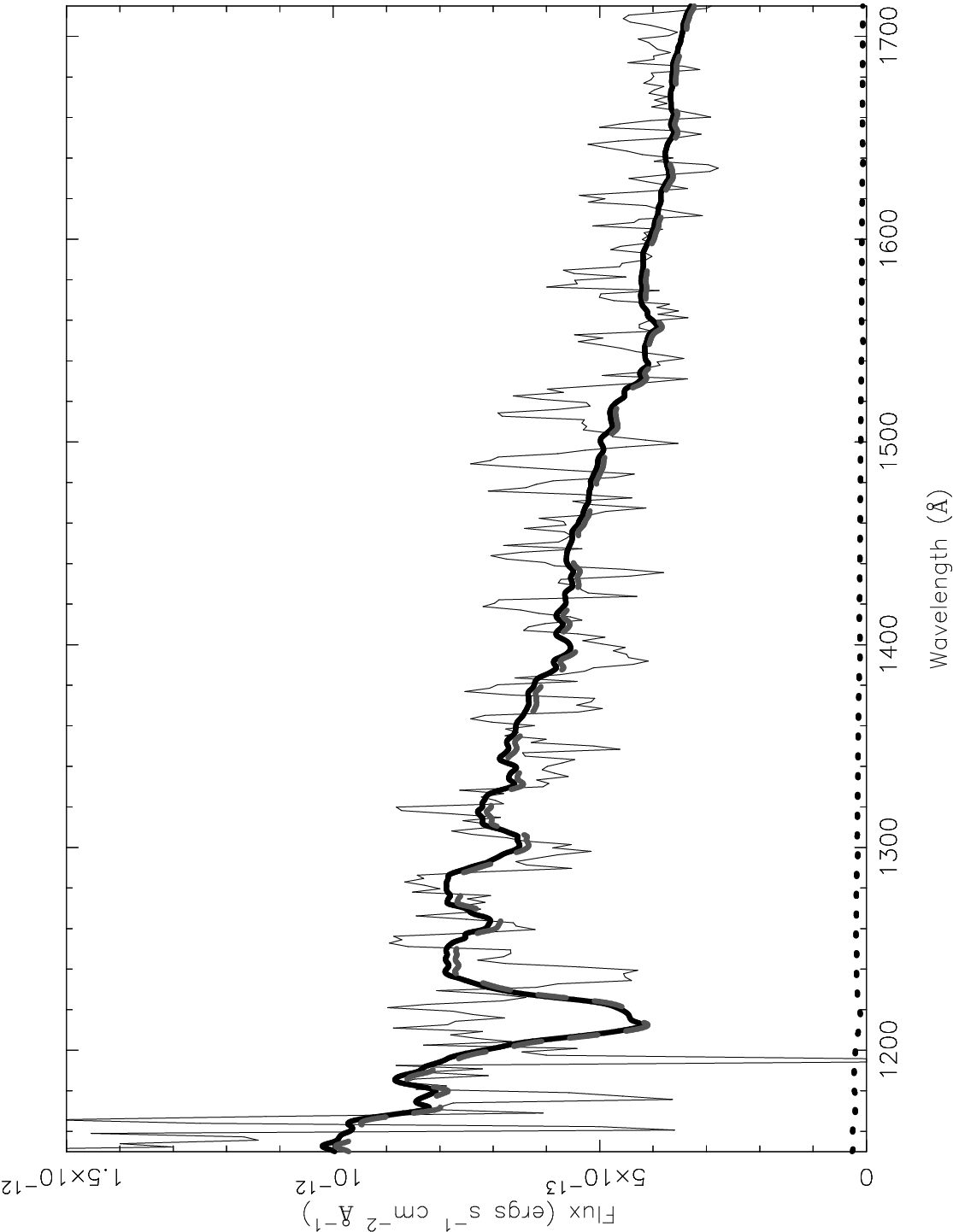}
%\clearpage

\plotone{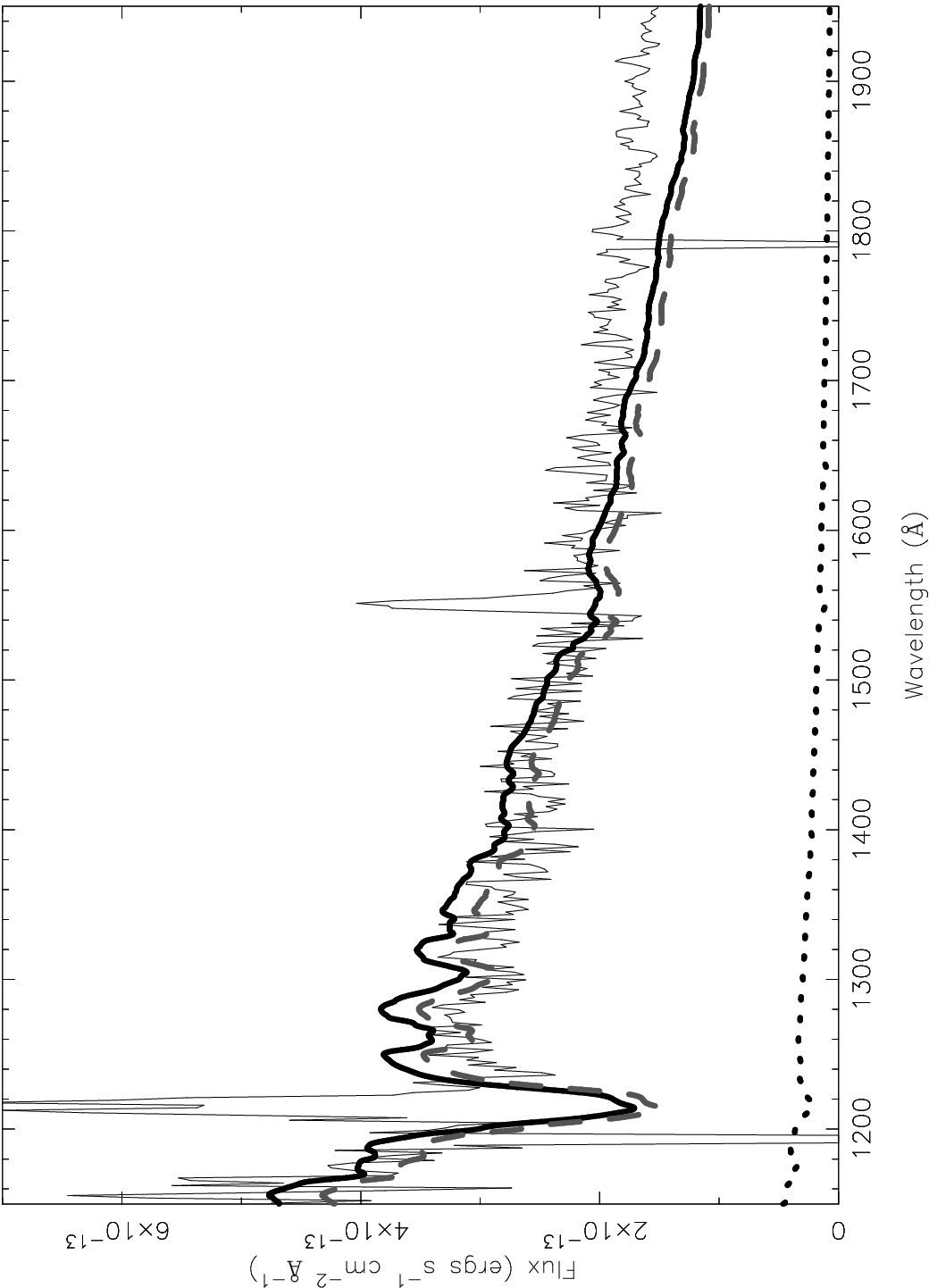}
%\clearpage

\plotone{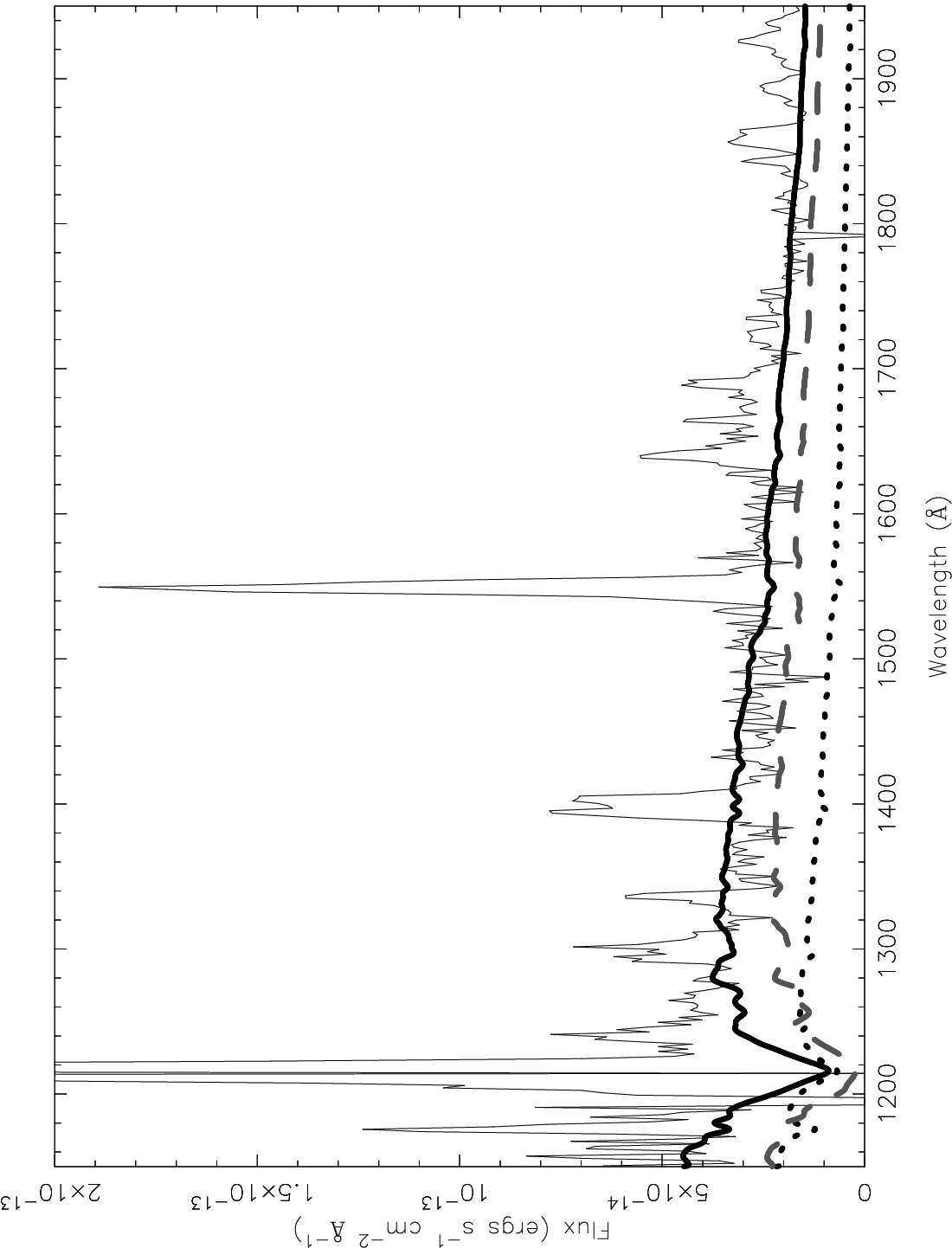}
\end{document}